\documentstyle[times,pramana,epsfig,floats]{ias}

\newcommand{\bm}{\bibitem}
\newcommand{\tauiso}{{\mbox{\boldmath $\tau$}}}
\newcommand{\gmu}{{\gamma_\mu}}

\begin{document}

\title{Strangeness production  
in proton-proton and proton-nucleus collisions{\footnote{Talk
presented in the workshop on Hadron Physics, Puri, India, March 7-17,2005}}}
\author{Radhey Shyam} 
\address{Saha Institute of Nuclear Physics, Kolkata, India 
 }
\maketitle \abstract{
In these lectures we discuss the investigation of the strange meson
production in proton-proton ($pp$) and in proton-nucleus ($pA$) reactions
within an effective Lagrangian model. The kaon production proceeds mainly 
via the excitations of $N^*$(1650), $N^*$(1710), and  $N^*$(1720) resonant 
intermediate nucleonic states, in the collision of two initial state
nucleons. Therefore, the strangeness production is expected to provide
information about the resonances lying at higher excitation energies. For
beam energies very close to the kaon production threshold the hyperon-proton
final state interaction effects are quite important. Thus, these studies
provide a check on the models of hyperon-nucleon interactions. The in-medium
production of kaons show strong sensitivity to the self energies of the
intermediate mesons.
}
%\vskip .3cm
%\maketitle 

\section{Introduction}

In the low energy domain Quantum Chromodynamics (QCD) is not amenable to 
the perturbation theory techniques. A compelling description of the QCD in 
this region could be achieved through Wilson's lattice guage theory
\cite{wil74}. These theories provide the most promising approach 
for the theoretical predictions of the properties of the hadronic ground
states and also of their excited states. However due to enormous computing
power necessary for the numerical treatment, lattice QCD has only started
to be able to describe baryon resonance masses and decay widths
\cite{zan02,sak02,lei03}. To simplify the problem a large number of
quark models for hadrons (see, eg.~\cite{the01}) have been
developed which aim at predicting the properties of hadrons by 
reducing the complexity of the strongly self-interacting multi-quark-gluon
system to an effective two- or three-quark system.

On the experimental side, the low energy behavior of QCD is mostly 
addressed indirectly. The nucleon is excited with the help of
a hadronic (nucleon or meson) or an electromagnetic (photon or
electron) probe. However, due to extremely short life time of the
excited hadron states, measurement of only their decay products is
possible. Consequently, experiments for investigating the inner 
structure of hadrons have focused on meson production off the nucleon.
Since, lattice QCD calculations are still far from being amiable to 
solutions for low and intermediate energy scattering reactions, it is
necessary to use effective methods for the description of the 
dynamical structure of these processes. Such effective methods account
for the inner structure of baryons by introducing explicit baryon 
resonance states, whose properties are extracted by comparison with
the experimental observables. The ultimate goal is to compare the
values extracted in this way with those predicted by the lattice 
QCD calculations.

In recent years, there has been a considerable amount of interest
in the study of the strangeness production reactions
in proton-proton ($pp$) collisions. This is expected to provide
information on the manifestation of QCD in the non-perturbative regime
of energies larger than those of the low energy pion physics where the
low energy theorem and partial conservation of axial current (PCAC)
constraints provide a useful insight into the relevant physics~\cite{eric88}.
The strangeness quantum number introduced by this reaction leads to new
degrees of freedom into this domain which are expected to probe the
admixture of $\bar{s}s$ quark pairs in the nucleon wave function~\cite{albe96}
and also the hyperon-nucleon and hyperon-strange meson interactions
\cite{delo89,adel90}.

The elementary nucleon-nucleon-strange meson production cross sections
are the most important ingredients in the transport model studies of 
the $K^+$-meson production in the nucleus-nucleus collisions, which
provide information on not only the initial collision dynamics but
also the nuclear equation of state at high
density~\cite{mosel91,brown91,maru94,misk94,hart94,liko94,liko95,liko98}.
Furthermore, the enhancement in the strangeness production 
has been proposed as a signature for the formation of the
quark-gluon plasma in high energy nucleus-nucleus
collisions~\cite{rafe82,knol88}. The understanding of the elementary
reactions are also a doorway to the investigation of the
production of hypernuclei in reactions like $A(p,K^+){_\Lambda}B$
where the hypernucleus ${_{\Lambda}}B$ has the
same neutron and proton numbers as the target nucleus $A$, with one
hyperon added. 

The measurements performed in late 1960s and 1970s provided the data on
the total cross sections for the associated hyperon ($Y$)-kaon production
at beam momenta larger than 2.80 GeV/c (these cross sections are
listed in Ref.~\cite{land88}).
With the advent of the high-duty proton-synchrotron (COSY) at the
Forschungszentrum, J\"ulich, it has become possible to perform
systematic studies of the associated strangeness production at beam
momenta very close to the reaction threshold (see, e.g., Ref.~\cite{mos02}
for a comprehensive review). At the near threshold beam energies, the
final state interaction (FSI) effects among the outgoing particles are
significant. Therefore, the new set of data are expected to probe also
the hyperon-nucleon and hyperon-strange meson interactions.

A very interesting result of the studies performed by the COSY-11
collaboration is that the ratio ($R$) of the total cross sections for
the $pp \to p\Lambda K^+$ and $pp \to p\Sigma^0 K^+$ reactions
(to be referred as $\Lambda K^+$ and $\Sigma^0 K^+$ reactions,
respectively) at the same excess energy
(defined as $\epsilon = \sqrt{s}-m_p-m_Y-m_K$, with $m_p$, $m_Y$, and
$m_K$ being the masses of proton, hyperon, and kaon respectively and $s$
the invariant mass of the collision), is about $28^{+6}_{-9}$ for
$\epsilon$ $<$ 13 MeV~\cite{sew99}. This result is very
intriguing because at higher beam energies ($\epsilon \approx$ 1000 MeV) 
this ratio is only around 2.5.

Several calculations have been reported \cite{gas00,sib00,lag01}
to explain this result. Assuming that the $\pi$- and $K$- exchange
processes are the only mechanism leading to the strangeness production,
the authors of Ref.~\cite{gas00} show within a (non-relativistic)
distorted wave Born approximation (DWBA) model that while the
$\Lambda K^+$ reaction is dominated by the $K$-exchange only, both
$K$- and $\pi$-exchange processes play an important role in the case of
$\Sigma^0 K^+$ reaction. Therefore, if the amplitudes
corresponding to the two exchanges in the latter case interfere
destructively, the production of $\Sigma^0$ is suppressed as compared
to that of $\Lambda$. It should however, be noted that in Ref.~\cite{sib00},
$K$- and $\pi$- exchange amplitudes are reported to be of similar
magnitudes for both $\Lambda K^+$ and $\Sigma^0K^+$ reactions.
In calculations reported in Ref.~\cite{lag01} also
the relative sign of $K-$ and $\pi-$ exchange terms is chosen solely 
by the criteria of reproducing the experimental data, although in this
work the theory has been applied to describe a wider range of data 
which includes the polarization transfer results of the DISTO
experiment~\cite{bal99} and the missing mass distribution obtained in
the inclusive $K^+$ production measurements performed at
SATURNE~\cite{sie94} apart from the ratio $R$.
Nevertheless, a conclusive evidence in support of the relative signs of
$\pi$- and $K$- exchange amplitudes being opposite to each other is still
lacking. Furthermore, other mechanisms like excitation, propagation, and
decay of intermediate baryonic resonances which play (see, e.g.,
\cite{shy99,shy01,col97}) an important role in the strangeness production
process, have not been considered by these authors. 
 
We have investigated the $\Lambda K^+$ and $\Sigma^0 K^+$ reactions at
near threshold as well as higher beam energies
in the framework of an effective Lagrangian approach (ELA)
\cite{shy99,shy01,shy96,shy98}. In this theory, the
initial interaction between two incoming nucleons is
modeled by an effective Lagrangian which is based on the exchange
of the $\pi$-, $\rho$-, $\omega$-, and $\sigma$- mesons. The coupling
constants at the nucleon-nucleon-meson vertices are determined by
directly fitting the T-matrices of the nucleon-nucleon ($NN$) scattering
in the relevant energy region. The ELA uses the pseudovector (PV) coupling
for the nucleon-nucleon-pion vertex which is consistent with the chiral
symmetry requirement of the quantum chromodynamics~\cite{wei68}. In
contrast to some earlier calculations~\cite{sib99}, both ($\Lambda K^+$
and $\Sigma^0K^+$) reactions proceed via excitation of the $N^*$(1650),
$N^*$(1710), and $N^*$(1720) intermediate baryonic resonance states.
The interference terms between the amplitudes of various resonances
are retained. To describe the near threshold data, the FSI effects in
the final channel are included within the framework of the Watson-Migdal
theory~\cite{wat52,shy98}. ELA has been used to describe
rather successfully the $pp \to pp\pi^0$, $pp \to pn\pi^+$
\cite{shy98,shy96}, $pp \to p K^+Y$ \cite{shy99,shy01} as well
as $pp \to ppe^+e^-$~\cite{shy03} reactions.

Within a similar approach we also investigate the $(p,K^+)$ reaction.
The initial interaction between the incoming proton and a bound
nucleon of the target is described by the one meson exchange 
processes. We use the same effective Lagrangians and vertex parameters to
model these interactions. The initial state interaction between
the two nucleons leads to the $N^*(1650)[\frac{1}{2}^-]$,
$N^*(1710)[\frac{1}{2}^+]$, and $N^*(1720)[\frac{3}{2}^+]$
baryonic resonance intermediate states. The vertex parameters
here too are the same as those used in the description of the elementary
reaction. 
 
\section{Effective Lagrangian Model}

The idea of the effective Lagrangian models is to account for the
symmetries of the QCD but including only effective degrees of 
freedom instead of quarks. These effective degrees of freedom are
modeled by baryons and mesons known to exist as (quasi-) bound
quark states. The advantage is that in this way one gets a better 
insight on the underlying production mechanism which makes the 
interpretation of the results easier. However, due to more 
complicated interaction structure, the meeting of the physical 
constraints like unitarity and analyticity becomes technically 
more involved. In fact, almost all the effective Lagrangian models
are not analytic, many of them are not even unitary. 

In our effective lagrangian model, we consider the tree-level structure
(Fig.~1) of the amplitudes for the associated $K^+ Y$ production in
proton-proton collisions, which  proceeds via the excitation of the
$N^*$(1650)($\frac{1}{2}^{-}$), $N^*$(1710)($\frac{1}{2}^{+}$), and
$N^*$(1720)($\frac{3}{2}^{+}$) intermediate resonances. The amplitudes
are calculated by a summation of the Feynman diagrams generated by means
of the effective Lagrangians at (a) the nucleon-nucleon-meson, (b) the
resonance-nucleon-meson, and (c) the resonance-$K^+$-hyperon vertices.
The assumption entering here is that the contributions of the higher-order
diagrams are negligible or can be absorbed in the form factors of the
first order diagrams.
\begin{figure}
\begin{center}
\epsfig{file=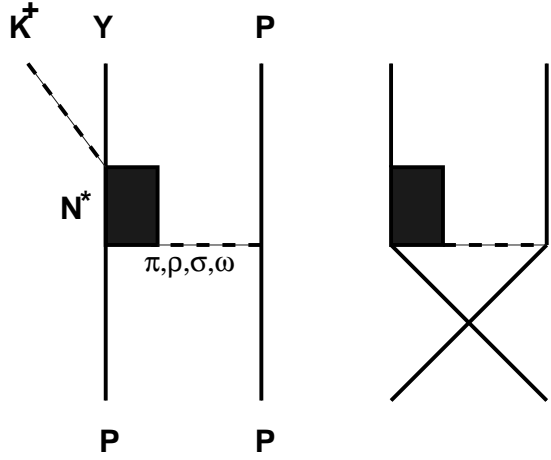,height=6cm}
\vskip 0.1in
\caption{  
Feynman diagrams for $K^{+}Y$ production in $pp$
collisions. The diagram on the left shows the direct process while that
on the right the exchange one.  }
\end{center}
\label{fig:figa}
\end{figure}
The parameters for $NN$ vertices are determined by fitting the $NN$
elastic scattering T matrix with an effective $NN$ interaction based
on the $\pi$, $\rho$, $\omega$ and $\sigma$ meson exchanges. The
effective meson-$NN$ Lagrangians are  
\begin{eqnarray}
{\cal L}_{NN\pi} & = & -\frac{g_{NN\pi}}{2m_N} {\bar{\Psi}}_N \gamma _5
                             {\gamma}_{\mu} \tauiso
                            \cdot (\partial ^\mu {\bf \Phi}_\pi) \Psi _N. \\
{\cal L}_{NN\rho} &=&- g_{NN\rho} \bar{\Psi}_N \left( \gmu + \frac{k_\rho}
                         {2 m_N} \sigma_{\mu\nu} \partial^\nu\right)
                          \tauiso \cdot \mbox{\boldmath $\rho$}^\mu \Psi_N. \\
{\cal L}_{NN\omega} &=&- g_{NN\omega} \bar{\Psi}_N \left( \gmu + \frac{k_\omega}
                         {2 m_N} \sigma_{\mu\nu} \partial^\nu\right)
                          \omega^\mu \Psi_N.   \\
{\cal L}_{NN\sigma} &=& g_{NN\sigma} \bar{\Psi}_N \sigma \Psi_N.
\end{eqnarray}
We have used the notations and conventions of Bjorken and Drell~\cite{bjor64}. 
In Eq.~(1) $m_N$ denotes the nucleon mass.  Note that we have used
a PV coupling for the $NN\pi$ vertex which is consistent with the chiral
symmetry requirement.
Since we use these Lagrangians to directly model the T-matrix, we have
also included a nucleon-nucleon-axial-vector-isovector vertex, with the
effective Lagrangian given by
\begin{eqnarray}
{\cal L}_{NNA} & = & g_{NNA} {\bar {\Psi}} \gamma_5 \gamma_\mu \tauiso \Psi
                     \cdot {\bf {A}}^\mu,
\end{eqnarray}
where $A$ represents the axial-vector meson field. This term is introduced
because in the limit of large axial meson mass ($m_A$) it cures the 
unphysical behavior in the angular distribution of $NN$ scattering caused by
the contact term in the one-pion exchange amplitude~\cite{sch94}, if
$g_{NNA}$ is chosen to be
\begin{eqnarray}
g_{NNA} =  \frac{1}{\sqrt{3}} m_A \left(\frac{f_\pi}{m_\pi}\right),
\end{eqnarray}
with very large ($\gg m_N$) $m_A$. $f_\pi$ appearing in Eq.~(6) is 
related to $g_{NN\pi}$ as $f_\pi = (\frac{g_{NN\pi}}{2m_N})m_\pi$.

We introduce, at each interaction vertex, the form factor
\begin{eqnarray}
F_{i}^{NN} & = & \left (\frac{\lambda_i^{2} - m_i^{2}}{\lambda_i^{2} - q_i^{2}}
        \right ), i= \pi, \rho, \sigma, \omega,
\end{eqnarray}
where $q_i$ and $m_i$ are the four momentum and mass of the $i$th 
exchanged meson, respectively. The form factors suppress the contributions of
high momenta and the parameter $\lambda_i$, which governs the
range of suppression, can be related to the hadron size. Since
$NN$ elastic scattering cross sections decrease gradually with the beam
energy (beyond certain value), we take energy dependent meson-nucleon
coupling constants of the following form 
\begin{eqnarray}
g(\sqrt{s}) & = & g_{0} exp(-\ell \sqrt{s}),
\end{eqnarray}
in order to reproduce these data in the entire range of beam energies. The
parameters, $g_0$, $\lambda$ and $\ell$ were determined by fitting
to the elastic proton-proton and proton-neutron scattering data at the
beam energies in the range of 400 MeV to 4.0 GeV~\cite{sch94,shy96}.
It may be noted that this procedure fixes also the signs of the
effective Lagrangians [Eqs.~(1)-(5)].
The values of various parameters are given in Table 1 of
Ref.~\cite{shy99}. The same parameters for these vertices were also
used in calculations of the pion and the dilepton production in  $pp$
collisions.  Thus we ensure that the $NN$ 
elastic scattering channel remains the same in the description of various
inelastic channels within this approach, as it should be. 
  
Below 2 GeV center of mass (c.m.) energy, only three resonances,
$N^*$(1650), $N^*$(1710), and $N^*$(1720), have significant decay
branching ratios into $KY$ channels. Therefore, we
have considered only these three resonances in our calculations.
The $N^*$(1700) resonance having very small (and uncertain) branching
ratio for the decay to these channels, has been excluded. 
Since all the three resonances can couple to the meson-nucleon channel 
considered in the previous section, we require the effective Lagrangians
for all the four resonance-nucleon-meson vertices corresponding to
all the included resonances. Since the mass of the strange quark is
much higher than that of the $u-$ or $d-$ quark, one does not expect
the pion like strict chiral constraints for the case of other
pseudoscalar mesons like $\eta$ and $K$ (to be called $\zeta$
in the following). Thus, one has a choice of psuedoscalar (PS) or PV
couplings for the $NN\zeta$ and $N^*_{1/2}N\zeta$ vertices (forms of the
corresponding effective Lagrangians are given in Ref.~\cite{shy99}).
The same holds also for the $N^*_{1/2}Y K$ vertices.

In principle, one can select a linear combination of both and fit the
PS/PV ratio to the data. However, to minimize the number of parameters
we choose either PS or PV coupling at a time. In the results shown below,
we have used PS couplings for both $N^*N\pi$ and $N^*\Lambda K^+$
vertices involving spin-1/2 resonances of even and odd parities.
Calculations have also been performed with the corresponding
PV couplings. The cross sections calculated with this option for
the $N_{1/2}^*YK$ vertex deviate very little from those
obtained with the corresponding PS couplings. However, data shows
a clear preference for the PS coupling at the $N_{1/2}^*N\pi$ vertices.

The couplings constants for the vertices involving resonances
are determined from the experimentally observed quantities such as
branching ratios for their decays to corresponding
channels. It may however, be noted that such a procedure can not be
used to determine the coupling constant for the $N^*(1650)\Sigma K$
vertices, as the on-shell decays of this resonance to 
$\Sigma K$ channel are inhibited. Therefore, we have tried to determine this
coupling constant by fitting the available data on the
$\pi^+p \to \Sigma^+ K^+$, $\pi^-p \to \Sigma^0 K^0$, and
$\pi^-p \to \Sigma^-K^+$ reactions in an effective Lagrangian  coupled
channels approach \cite{feu98,pen02}, where all the available data for the
transitions from $\pi N$ to five meson-baryon final states, $\pi N$,
$\pi \pi N$, $\eta N$, $K\Lambda$, and $K\Sigma$ are simultaneously
analyzed for center of mass energies ranging from threshold to 2 GeV. In
this analysis all the baryonic resonances with spin $\leq \frac{3}{2}$
up to excitation energies of 2 GeV are included as intermediate states. 
Since the resonances considered in this study have no known
branching ratios for the decay into the $N\omega$ channel, we determine the
coupling constants for the $N^*N\omega$ vertices by the strict
vector meson dominance (VMD) hypothesis~\cite{saku69}, which  
is based essentially on the assumption that the coupling of photons on hadrons 
takes place through a vector meson.  

It should be stressed that the branching ratios determine only
the square of the corresponding coupling constants; thus their signs remain
uncertain in this method. Predictions from independent calculations
(${\it e.g}$ the quark model) can, however, be used to constrain these
signs. The magnitude as well as signs of the coupling constants for the
$N^*N\pi$, $N^*\Lambda K$, $N^*N\rho$, and $N^*N(\pi \pi)_{s-wave}$
vertices were determined by Feuster and Mosel~\cite{feu98} and Manley
and Saleski~\cite{manl92} in their analysis of the pion-nucleon data
involving the final states $\pi N$, $\pi \pi N$, $\eta N$, and $K\Lambda$.
Predictions for some of these quantities are also given in the
constituent quark model calculations of Capstick and Roberts~\cite{caps94}.
Guided by the results of these studies, we have chosen the positive
sign for the coupling constants for these vertices. Unfortunately,
the quark model calculations for the $N^*N\omega$ vertices are still
sparse and an unambiguous prediction for the signs of the corresponding
coupling constants may not be possible at this stage~\cite{stan93}.
Nevertheless, we have chosen a positive sign for the coupling constants
for these vertices as well. Values of all the coupling constants are
given in Ref.~\cite{shy99}.

After having established the effective Lagrangians,
coupling constants and form of the propagators (which are given in 
Ref.~\cite{shy99}), it is straight forward to 
write down the amplitudes for various diagrams associated with
the $pp \to pYK$ reactions which can be calculated numerically by following
${\it e.g.}$ the techniques discussed in~\cite{shy96}. The isospin
part is treated separately. This
gives rise to a constant factor for each graph, which is unity for the
reaction under study. It should be noted that the signs of
various amplitudes are fixed by those of the effective Lagrangian
densities, coupling constants and propagators
as described above. These signs are not allowed to change anywhere in
the calculations.

In the present form of our effective Lagrangian theory, the 
energy dependence of the cross section due to FSI is separated from
that of the primary production amplitude and the total amplitude is
written as,
\begin{eqnarray}
A_{fi} & = & M_{fi}(pp \rightarrow pYK^+) \cdot T_{ff},
\end{eqnarray}
where $M_{fi}(pp \rightarrow pYK^+)$ is the primary associated 
$YK$ production amplitude, while $T_{ff}$ describes the re-scattering
among the final particles which goes to unity in the limit of no FSI.
The latter is taken to be the coherent sum of the two-body on-mass-shell
elastic scattering amplitudes $t_i$ (with $i$ going from 1 to 3),
of the interacting particle pairs $j-k$ in the final
channel. This type of approach has been used earlier to
describe the pion~\cite{shy98,dub86,mei98}, $\eta$-meson
\cite{mol96,dru97,del04}, $\Lambda K^+$ \cite{shy99} and $\phi$-meson
\cite{tit00} production in $pp$ collisions.
\begin{figure}[ht]
\centerline{\epsfig{file=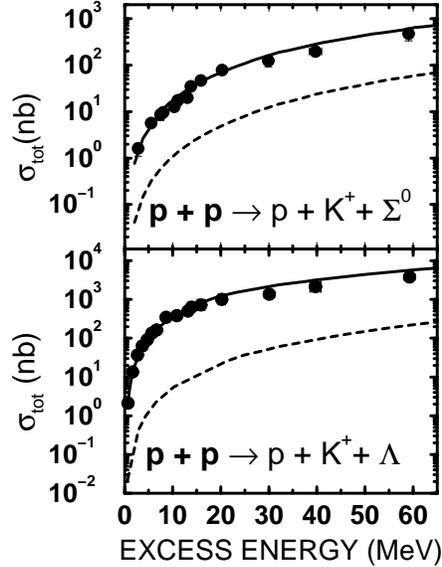,height=8cm}}
\caption{  
Comparison of the calculated and the experimental total
cross sections for the $pp \to p\Lambda K^+$ 
and $pp \to p\Sigma^0K^+$ reactions as a function of
the excess energy. Results obtained with no FSI effects
are shown by dashed lines. The experimental data are from
Refs.~\protect\cite{sew99}
  }.
\label{fig:figb}
\end{figure}
An assumption inherent in Eq.~(9) is that the reaction takes place over
a small region of space (which is fulfilled rather well in
near threshold reactions involving heavy mesons). Under this condition the
amplitudes $t_i$ can be expressed in terms of the inverse of the
Jost function~\cite{wat52,shy98} which has been calculated by
using a Coulomb modified effective range expansion of the
phase-shift~\cite{noy72}. The required effective range and
scattering length parameters are given in Refs.~\cite{shy99,shy01}.  

\section{Kaon production in proton-proton collisions}

The total cross sections for the $\Lambda K^+$ and $\Sigma^0 K^+$
reactions as a function of the excess energy are shown in Fig.~2.
The calculations are the coherent sum of all resonance excitation and meson
exchange processes as described earlier. In both cases a good agreement is 
obtained  between theory and the data available from the COSY-11
collaboration. Keeping in mind the fact that all parameters of the
model, except for those of $N^*Yp$ vertices and the FSI, were the
same in the two calculations and that no parameter was freely varied,
this agreement is quite satisfactory. It should be noted that we do not
require to introduce arbitrary normalization constants to get the agreement
between calculations and the data.  We also show in this figure the
results obtained without including the FSI effects (dashed line). It
can be seen that the FSI effects are vital for a proper description of 
the experimental data in both the cases.

In Fig.~3, we have investigated the role of various meson exchange
processes in describing the total cross sections. The dashed, long-dashed,
dashed-dotted, and solid with black square curves represent the
contributions of $\pi$, $\rho$, $\omega$ and $\sigma$ meson exchanges,
respectively. The contribution of the heavy axial meson exchange
is not shown in this figure as it is negligibly small. The coherent
sum of all the meson-exchange processes is shown by the solid line.
It is clear that the pion exchange graphs dominate the production
process for both the reactions in the entire range of beam energies.
Contributions of $\rho$ and $\omega$ meson exchanges are almost
insignificant. On the other hand, the $\sigma$ meson exchange, which
models the correlated $s-$wave two-pion exchange process 
and provides about 2/3 of this exchange in the low energy $NN$
interaction, plays a relatively more important role. This observation
has also been made in case of the $NN\rightarrow NN\pi$
reaction~\cite{dmit86,risk93,horo94,shy96}. 
\begin{figure}[ht]
\centerline{\epsfig{file=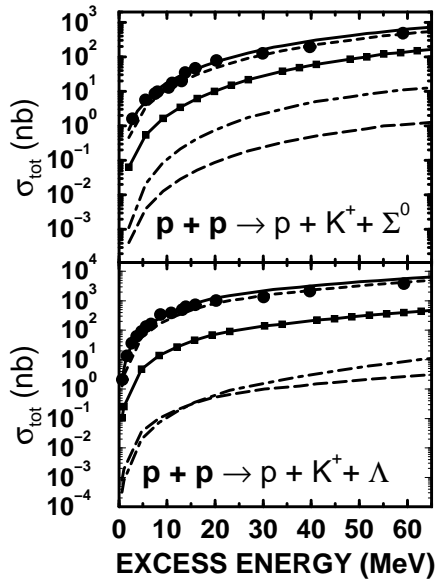,height=8cm}}
\caption{  
Contributions of various meson exchange processes to the
total cross section for the $pp \rightarrow pK^{+}\Lambda$
and $pp \rightarrow pK^{+}\Sigma^0$ reactions. The dashed, long-dashed,
dashed-dotted and solid with black squares curves represent the
contributions of $\pi$, $\rho$, $\omega$ and $\sigma$ meson exchanges,
respectively. Their coherent sums are shown by the solid lines }.
\label{fig:figc}
\end{figure}

The individual contributions of various nucleon resonances to the
total cross sections of the two reactions are shown in Fig.~4. 
We note that in both
the cases, the cross section is dominated by the $N^*$(1650)
resonance excitation for $\epsilon < 30$ MeV. Since $N^*$(1650) is
the lowest energy baryonic resonance having branching
ratios for the decay to $YK^+$ channels, its dominance in
these reactions at beam energies very close the kaon production threshold
is to be expected. In the near threshold region the relative dominance
of various resonances is determined by the dynamics of the reaction
where the difference of about 60 MeV in excitation energies of
$N^*$(1650) and $N^*$(1710) resonances plays a crucial role.
\begin{figure}[ht]
\centerline{\epsfig{file=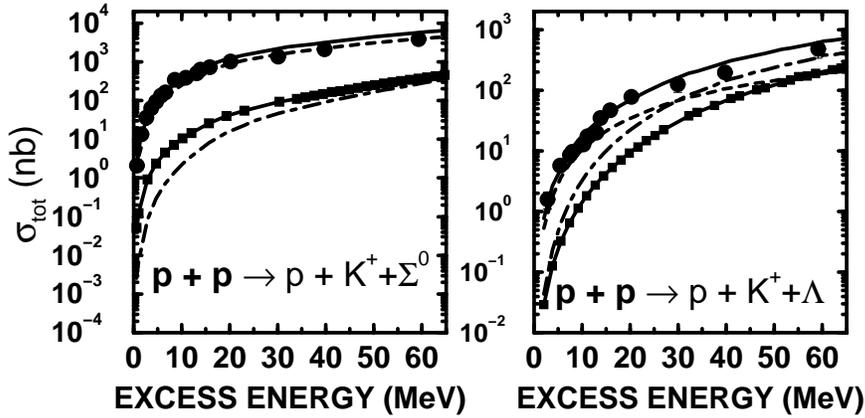,height=10cm}}
\caption{  
Contributions of N$^{*}$(1650) (dashed line), N$^{*}$(1710)
(full line with black squares) and N$^{*}$(1720) (dashed-dotted line)
baryonic resonances to the total cross section for the two reactions
studied in Fig.~2. Their coherent sum is shown by the solid line.}
\label{fig:figd}
\end{figure}
However, for $\epsilon$ values between 30 - 60 MeV, while the
$pp \to pK^+\Lambda$ reaction continues to be dominated by $N^*(1650)$
excitation, the $pp \to pK^+\Sigma^0$ reaction gets significant
contributions also from $N^*(1710)$ and $N^*(1720)$ resonances. This
difference in the role of the three resonances in the two cases can be
understood in the following way. For a resonances to contribute
significantly, we should have  $m_Y + m_K + \epsilon \geq m_R +
\Gamma_R/2$, where $m_R$ and $\Gamma_R$ are the mass and width of the
resonance, respectively. Therefore, in the region of excess energies
$\geq Q [ = (m_R + \Gamma_R/2) - (m_Y + m_K)]$, the particular resonance
$R$ should contribute significantly. The values of $Q$ for the $pp
\to pK^+\Lambda$ reaction, are 115 MeV, 150 MeV, and 185 MeV, for
the $N^*(1650)$, $N^*(1710)$, and $N^*(1720)$ resonances, respectively.
On the other hand, for the $pp \to p K^+\Sigma^0$ case, they are 36 MeV,
72 MeV and 105 MeV, respectively for these three resonances. Therefore,
contributions of the $N^*(1710)$ and $N^*(1720)$ resonance excitations are
negligibly small for the $K^+\Lambda$ production in the entire energy
range of the COSY-11 data (i.e., for $\epsilon \leq$ 60. MeV) while they are
significant for the $K^+\Sigma^0$ case for $\epsilon >$ 30 MeV. It would
be helpful to have data on the invariant mass spectrum at these excess
energies in order to  confirm these theoretical observations.

In Fig.~5, we compare our calculations with the data
for the ratio $R$ as a function of $\epsilon$. We have shown here the
results for excess energies up to 60 MeV, where the COSY-11 data are 
available. It is clear that our calculations are able to describe well 
the trend of the fall-off of $R$ as a function of the excess energy. 
It should be noted that  FSI effects account for about $60\% - 80\%$
of the observed ratio for $\epsilon < 30$ MeV and about 
50$\%$ of it beyond this energy. Therefore, not all of the observed value of
$R$ at these beam energies can be accounted for by the FSI effects, which
is in agreement with the observation made in \cite {gas00}. It should
again be emphasized that without considering the contributions of the
$N^*$(1650) resonance for the $\Sigma^0K^+$ reactions, the
calculated ratio would be at least an order of magnitude larger.
Therefore, these data are indeed sensitive to the details of the
reaction mechanism. At higher beam energies ($\epsilon$ $>$ 300 MeV),
values of $R$ obtained with and without FSI effects are almost identical.
In this region the reaction mechanism is different; here all the three
resonances contribute in one way or the other, their
interference terms are significant~\cite{shy99}, and FSI related
effects are unimportant.
\begin{figure}[ht]
\centerline{\epsfig{file=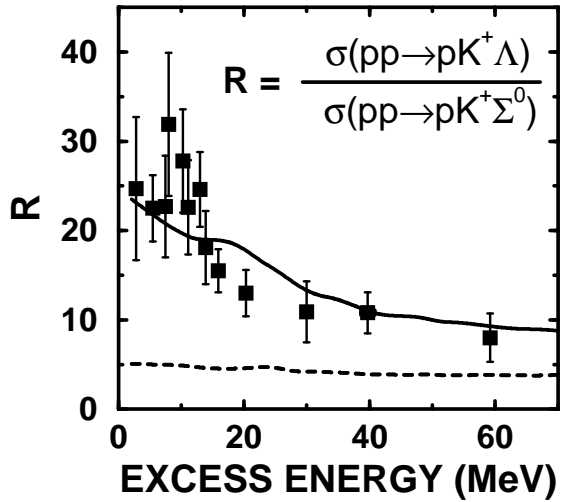,height=10cm}}
\caption{  
Ratio of the total cross sections for $pp \to p\Lambda K^+$ and
$pp \to p\Sigma^0 K^+$ reaction as a function
of the excess energy. The solid and dashed lines show the results of our
calculations with and without FSI effects, respectively. The data are from
Refs.~\protect\cite{sew99}.
 }
\label{fig:fige}
\end{figure}

\section{Kaon production in proton-nucleus collisions}

The study of meson production in proton-nucleus collisions is an 
interesting tool to investigate the influence of nuclear medium 
on the properties of hadrons and their production processes. Since
the mass of kaon is larger than that of the pion, medium effects 
play even more vital role in its production in $pA$ collisions.

The kaon production reactions of the type $p + A(N,Z) \rightarrow
{_{\Lambda}}B(N,Z) + K^+$ (where $N$ and $Z$ are the neutron and proton
numbers, respectively, in the target the target nucleus $A$), leads
to the production of the $\Lambda$-hypernucleus ${_{\Lambda}}B(N,Z)$.
The study of this reaction, therefore, is likely to lead to an 
understanding of the hypernuclear spectroscopy. At the same time,
since this reaction involves a large momentum transfer to the 
nucleus, it provides an appropriate tool to learn about the 
behaviour of the nuclear many body wave function at higher momenta
which is not very well known. We present here some results of our
investigation for this reaction which will be referred to as
$A(p,K^+){_{\Lambda}}B$ reaction. 

The model used to describe the $(p,K^+)$ reaction is similar to 
the effective Lagrangian approach described above [see, Figs.~(6b) and
(6c)]. The initial interaction between the incoming proton and a bound
nucleon of the target is described by the one-meson exchange mechanism.
We use the same effective Lagrangians and vertex parameters to model these
interactions.  The initial state interaction between the two nucleons leads
to the $N^*(1650)[\frac{1}{2}^-]$, $N^*(1710)[\frac{1}{2}^+]$, and
$N^*(1720)[\frac{3}{2}^+]$ baryonic resonance intermediate states. The
vertex parameters here too are the same as those used in the description
of the elementary reactions.  
\begin{figure}[ht]
\centerline{\epsfig{file=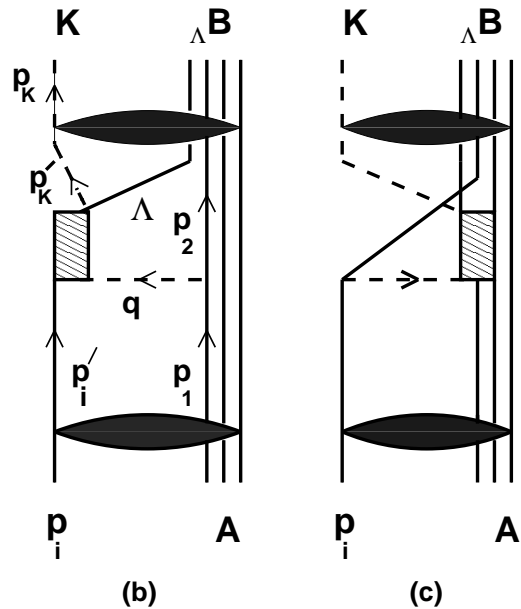,height=8cm}}
\vskip 0.1in
\caption{Graphical representation of the  $A(p,K^+){_{\Lambda}}B$ reaction
models. The elliptic shaded area represent the optical
model interactions in the incoming and outgoing channels.
}
\label{fig:figf)}
\end{figure}
It may be noted that in this model there are altogether three active
bound state baryon wave functions taking part in the reaction process
allowing the large momentum transfer to be shared among three baryons.
Consequently, the sensitivity of the model is shifted from high momentum
parts of the bound state wave functions (not very well known)
to those at relatively lower momenta where they are rather
well known from the $(e, e^\prime p)$ and $(\gamma, p)$ experiments
(see, e.g.,~\cite{fru84}) and
are relatively larger. This type of two-nucleon model has recently 
been applied to the study the $A(p,K^+){_{\Lambda}}B$ reaction in
Ref.~\cite{shy04,shy05} where to reduce the computational
complications plane waves have been used to describe the
relative motions of incoming proton and outgoing kaon.

In performing calculations for the cross sections of the
$A(p,K^+){_{\Lambda}}B$ reactions, one requires spinors for the final
bound hypernuclear state (corresponding to momentum $p_\Lambda$) and for
two intermediate nucleonic states (corresponding to momenta $p_1$ and $p_2$)
These are determined by assuming them to be pure-single particle or
single-hole states with the core remaining inert. The quantum numbers of
the two intermediate states are taken to be the same. The spinors in the
momentum space are obtained by Fourier transformation of the
corresponding coordinate space spinors which are obtained by solving
the Dirac equation with potential fields consisting of an attractive
scalar part  and a vector part with the Woods-Saxon geometry. With a
fixed set of the geometry parameters the depths of the potentials are
searched in order to reproduce the binding energies of the particular state
\cite{shy05}.

To have an idea of the relative strengths of the upper and lower
components of the Dirac spinors as a function of the transferred momentum,
we show, e.g., in Fig.~7 the $0p_{3/2}$ $\Lambda$ hyperon spinors in
momentum space for the $^{41}\!\!\!_\Lambda$Ca hypernucleus.  We
note that only for momenta $<$ 1.5 fm$^{-1}$, is the lower component of
the spinor substantially smaller than the upper component. In the region
of momentum transfer pertinent to to exclusive kaon production in
proton-nucleus collisions, the magnitudes of the upper and lower components
are of the same order of magnitude. This clearly demonstrates that a
fully relativistic approach is essential for an accurate description of
this reaction.
\begin{figure}[ht]
\centerline{\epsfig{file=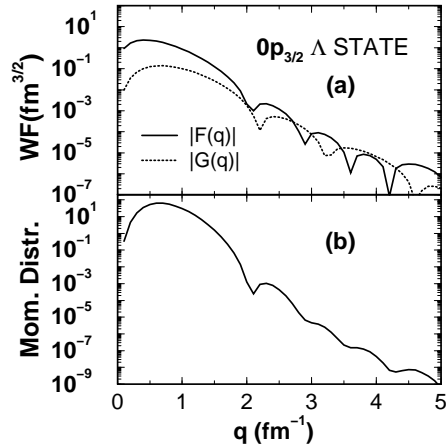,height=6cm}}
\caption{ (a) Momentum space spinors (WF) for $0p_{3/2}$ $\Lambda$
orbit in $^{41}\!\!\!_\Lambda$Ca hypernucleus. $|F(q)|$ and $|G(q)|$ are
the upper and lower components of the spinor, respectively. 
(b) Momentum distribution (Mom. Distr.) for the same state calculted
with wave functions shown in (a).
}
\label{fig:figg)}
\end{figure}

The self-energies of the exchange mesons are another input quantities
required in the calculations of the $A(p,K^+){_{\Lambda}}B$ reaction.
They take into account the medium effects on the intermediate meson
propagation. The $\rho$ and $\omega$ self energies have been calculated
by following the procedure described in Ref.~\cite{muh04}. 
The pion self energy is more crucial as one-pion exchange diagrams
dominate the $(p,K^+)$ cross sections. This is obtained by calculating
the contribution of particle-hole ($ph$) and delta-hole ($\Delta h$)
excitations produced by the propagating pion~\cite{dmi85}. This has been
renormalized by including the short-range repulsion effects by introducing
the constant Landau-Migdal parameter $g^\prime$ which is taken to be the
same for $ph-ph$ and $\Delta h-ph$ and $\Delta h-\Delta h$ correlations
which is a common choice. The parameter $g^\prime$, acting in the
spin-isospin channel, is supposed to mock up the complicated density
dependent effective interaction between particles and holes in the nuclear
medium. Most estimates give a value of $g^\prime$ between 0.5 - 0.7. 
The sensitivity of the pion self-energy [$\Pi(q)$]to the $g^\prime$
parameter is studied in Ref.~\cite{shy04}.   
\begin{figure}
\centerline{\epsfig{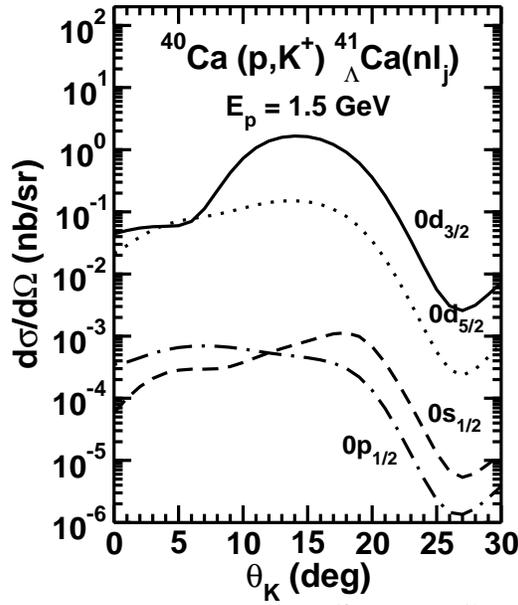}}
\caption{Differential cross section for the
$^{40}$Ca$(p,K^+)$$^{41}\!\!\!_\Lambda$Ca
reaction for the incident proton energy of 1.5 GeV for various
bound states of final hypernucleus as indicated in the figure.
The $\Lambda$ separation energies for $0d_{3/2}$, $0d_{5/2}$, 
$0p_{1/2}$, and $0s_{1/2}$ states were taken to be 0.7529 MeV,
1.5435 MeV, 9.1400 MeV, and 17.8802 MeV, respectively. The quantum
number and the binding energy of the two intermediate nucleon states
were $0d_{3/2}$ and 8.3282 MeV, respectively.
}
\label{fig:figh)}
\end{figure}

In Fig.~8, we show the kaon angular distributions corresponding to
various final hypernuclear states excited in the reaction
$^{40}$Ca$(p,K^+)$$^{41}\!\!\!_\Lambda$Ca.
We have taken $g^\prime = 0.5$ throughout in this figure. It may be noted
that in all the cases the diagram 6(b) makes a dominant contribution to
the cross sections.  Clearly, the cross sections are quite selective about
the excited hypernuclear state, being maximum for the state of largest orbital
angular momentum. This is due to the large momentum transfer involved
in this reaction. We see in this figure that in each case the angular
distribution has a maximum at angles larger than the zero degree and not
at the 0$^\circ$ as seen in previous non-relativistic calculations of  
this reaction. This is the consequence of using Dirac spinors for
the bound states. There are several maxima in the upper and lower
components of the momentum space bound spinors in the
region of large momentum transfers. Therefore, in the kaon angular
distribution the first maximum may shift to larger angles.

In Fig.~9, we show the dependence of our calculated cross sections on
pion self-energy. It is interesting to note that the self energy has a rather
large effect. We also see a surprisingly large effect on the
short range correlation (expressed schematically by the Landau- Migdal
parameter $g^\prime$) on the cross sections. 
                                                                                
The absolute magnitudes of the cross sections near the peak is around
1-2 nb/sr, although the distortion effects could reduce these values
as is shown~\cite{shy05}. This order of magnitude estimates should be
useful in planning of the future experiments for this reaction. As found
in Ref.~\cite{shy04}, contributions from the $N^*$(1710) resonance dominate
the total cross section in each case. Also the interference
terms of the amplitudes corresponding to various resonances are not
negligible. It should be emphasized that we have no freedom in choosing
the relative signs of the interference terms.
\begin{figure}
\centerline{\epsfig{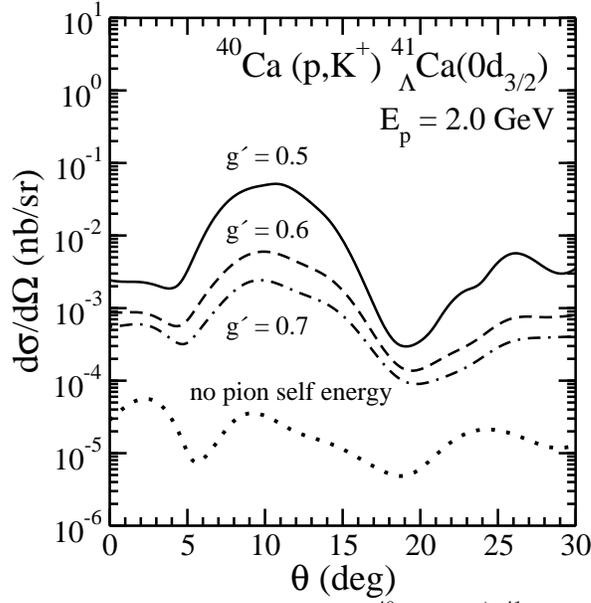}}
\caption{Differential cross section for the
$^{40}$Ca$(p,K^+)$$^{41}\!\!\!_\Lambda$Ca($0d_{3/2}$)
reaction for the incident proton energy of 2.0 GeV.
The dotted line shows the results obtained without including the pion
self-energy in the denominator of the pion propagator while
full, dashed and dashed-dotted lines represent the same calculated
with pion self-energy renormalized with Landau-Migdal parameter of
0.5,0.6 and 0.7, respectively.
}
\label{fig:figi)}
\end{figure}

\section{Summary and conclusions}

In summary, our effective lagrangian model  
can describe well the recent data on $pp \to p\Lambda K^+$ and
$pp \to p\Sigma^0K^+$ reactions. An important result is that
the $N^*$(1650) resonant state contributes predominantly to both these
reactions at near threshold beam energies. Therefore, these reactions in
this energy regime, provide an ideal means of investigating the properties
of this $S_{11}$ baryonic resonance. To the extent that the final state
interaction effects in the exit channel can be accounted for by the
Watson-Migdal theory, our model is able to explain the experimentally
observed large ratio of the total cross sections of the two reactions in
the near threshold region.

We also investgated the $A(p,K^+){_\Lambda}B$ reaction within a 
similar approach. We find that the nuclear medium corrections to the
intermediate pion propagator introduce large effects on the kaon differential
cross sections. There is also the sensitivity of the cross sections
to the short-range correlation parameter $g^\prime$ in the pion self-energy.
Thus, $(p,K^+)$ reactions may provide an interesting alternative
tool to investigate the medium corrections on the pion propagation
in nuclei. Moreover, the study of the $(p,K^+)$ reaction is
attractive as it provides a  way to study the
the spectroscopy of the $\Lambda$ hypernuclear states. 
This reaction should be measurable at the COSY facility in the
Forschungszentrum J\"ulich. The characteristics of these cross sections
predicted by us should be helpful in planning of such experiments. 

\section{Acknowledgements}
The author wishes to acknowledge useful discussions with Bo H\"oistad, 
G\"oran F\"aldt, Anders Ingemarsson, Horst Lenske, Ulrich Mosel, 
Walter Oelert, and Gregor Penner.


\begin{thebibliography}{99}
\bm{wil74}
K. G. Wilson, Phys. Rev. D {\bf 10} (1974) 2445.

\bm{zan02}
J. M. Zanotti, S. Bilson-Thompson, F. D. R. Bonnet, P. D. Coddington,
D. B. Leinweber, A. G. Williams, J. B. Zhang, W. Lelnitchouk, and 
F. X. Lee, Phys. Rev. D {\bf 65} (2002) 074507.

\bm{sak02}
S. Sakai, T. Blum, and S. Ohta, Phys. REv. D {\bf 65} (2002) 074503.

\bm{lei03}
D. B. Leinweber, A. W. Thomas, A. G. Williams, R. D. Young, J. M. Zanotti,
and J. B. Zhang, Nucl. Phys. A {\bf 737} (2004) 177.

\bm{the01}
L. Theussl and R. F. Wagenbrunn, Phys. Rev. C {\bf 64} (2001) 068201:
L. Theussl, R. F. Wagenbrunn, B. Desplanques, and W. Plessas, Eur. Phys. J.
{\bf A12} (2001)91, Nucl. Phys. {\bf A689} (2001) 394.

\bm{eric88}
T.E.O Ericson and W. Weise, ${\it Pions \, and \, Nuclei}$,
Clarendon, Oxford, 1988.

\bm{albe96}
M. Alberg, Prog. Part. Nucl. Phys. {\bf 36} (1996) 217.

\bm{delo89}
A. Deloff, Nucl. Phys. {\bf A505} (1989) 583.

\bm{adel90}
R.A. Adelseck and B. Saghai, Phys. Rev. C {\bf 42} (1990) 108.

\bm{mosel91}
U. Mosel, Annu. Rev. Nucl. Part. Sci. {\bf 41} (1991) 29 and
references therein.

\bm{brown91}
G.E. Brown, C.M. Ko, Z.G. Wu and L.H. Xia, Phys. Rev.C {\bf 43}
(1991) 1881.

\bm{maru94}
T. Maruyama, W. Cassing, U.Mosel, S. Teis and K. Weber,
Nucl. Phys. {\bf A573} (1994) 653.

\bm{misk94}
D. Miskowiec et al., Phys. Rev. Lett. {\bf 72} (1994) 3650.

\bm{hart94}
C. Hartnack, J. Jaenicke, L. Sehn, H. Str\"ocker and
J. Aichelin, Nucl. Phys. {\bf A580} (1994) 643.

\bm{liko94}
X.S. Fang, C.M. Ko, G.Q. Li, Y.M. Zheng, Nucl. Phys. {\bf A575}
(1994) 766.

\bm{liko95}
G.Q. Li and C.M. Ko, Phys. Lett. {\bf B349} (1995) 405.

\bm{liko98}
G.Q. Li, C.M. Ko and W.S. Chung, Phys. Rev.C {\bf 57} (1998) 434.

\bm{rafe82}
J. Rafelski and B. M\"uller, Phys. Rev. Lett. {\bf 48} (1982) 1066.

\bm{knol88}
H.W. Barz, B.L. Friman, J. Knoll and H. Schulz, Nucl. Phys. {\bf A485}
(1988) 685.

\bm{land88}
Landolt-B\"ornstein, New Series, Ed. H. Schopper, I/12 (1988).

\bm{mos02}
P. Moskal, M. Wolke, A. Khoukaz, and W. Oelert, Prog. Part. Nucl. Phys.
{\bf 49} (2002) 1.
 
\bm{sew99}
S. Sewerin et al., Phys. Rev. Lett. {\bf 83}, 682 (1999);
P. Kowina et. al., Eur. Phys. J {\bf A 22} (2004) 293. 

\bm{gas00}
A.M. Gasparian, J. Haidenbauer, C. Hanhart, L. Kondratyuk, and
J. Speth, Phys. Lett. B {\bf 480}, 273 (2000).

\bm{sib00}
A. Sibirtsev, K. Tsushima, W. Cassing, and A.W. Thomas, LANL
preprint, nucl-th/0004022.

\bm{lag01}J. M. Laget, Nucl. Phys. {\bf A691} (2001) 11c.; {\it ibid.},
Phys. Lett. {\bf B359} (1991) 24.

\bm{shy99}
R. Shyam, Phys. Rev. C {\bf 60} (1999) 055213.

\bm{shy01}
R. Shyam, G. Penner, and U. Mosel, Phys. Rev. C {\bf 63} (2001) 022202(R).

\bm{col97}
G. F\"aldt and C. Wilkin, Z. Phys. A {\bf 357} (1997) 241.

\bm{bal99}
F. Balestra et al., Phys. Rev. Lett. {\bf 83} (1999) 1534.

\bm{sie94}
R. Siebert et al., Nucl. Phys. {\bf A567} (1994)819.

\bm{sib99}
A. Sibirtsev, K. Tsushima, and A.W. Thomas, Phys. Lett. {\bf B421} (1998) 59;
A. Sibirtsev, K. Tsushima, W. Cassing, and A.W. Thomas, Nucl. Phys. {\bf A646}
(1999) 427. 

\bm{wat52}
M.L. Goldberger and K.M. Watson, {\it Collision Theory}, Wiley,
New York, 1969, pp 549.

\bm{shy96}
A. Engel, R. Shyam, U. Mosel and A.K. Dutt-Majumdaer, Nucl. Phys.A
{\bf 603} (1996) 387.

\bm{shy98}
R. Shyam and U. Mosel, Phys. Lett. B {\bf 425} (1998) 1.

\bm{shy03}
R. Shyam and U. Mosel, Phys. Rev. C {\bf 67} (2003) 065202.

\bm{wei68}
S. Weinberg, Phys. Rev. {\bf 166} (1996) 1568.

\bm{bjor64}
J.D. Bjorken and S.D. Drell, {\it Relativistic Quantum Mechanics},
McGraw-Hill, New York, 1964.

\bm{sch94}
M. Sch\"afer, H.C. D\"onges, A. Engel and U. Mosel,
Nucl. Phys.A {\bf 575} (1994) 429.

\bm{feu98}
T. Feuster and U. Mosel, Phys. Rev. C {\bf 58} (1998) 457;
{\it ibid.} {\bf 59} (1999) 460.

\bm{pen02}
G. Penner and U. Mosel, Phys. Rev. C {\bf 66} (2002) 055211; {\bf 66}
(2002) 055212.

\bm{saku69}
J.J. Sakurai, {\it Currents and Mesons}, Univ. of Chicago Press,
Chicago, 1969; Ann. Phys. {\bf 11} (1960) 1.

\bm{manl92}
D.M. Manley and E.M. Saleski, Phys. Rev.D {\bf 45} (1992) 4002. 

\bm{caps94}
S. Capstick and W. Roberts, Phys. Rev.D {\bf 49} (1994) 4570.

\bm{stan93}
Fl. Stancu and P. Stassart, Phys. Rev.D {\bf 46} (1993) 2140.

\bm{dub86}
J. Dubach, W.M. Kloet, and R.R. Silbar, Phys. Rev. C {\bf 33}
(1986) 373.

\bm{mei98}
V. Bernard, N. Kaiser, and Ulf-G. Meissner, Eur. Phys. J. A
{\bf 4} (1999) 259.

\bm{mol96}
A. Moalem, E. Gedalin, L. Razdolskaja, and Z. Shorer,
Nucl. Phys. A {\bf 600} (1996) 445.

\bm{dru97}
B.L. Druzhinin, A.E. Kudryavtsev, and V.E. Tarasev, Z. Phys. A
{\bf 359} (1997) 205.

\bm{del04}
A. Deloff, Phys. Rev. C {\bf 69} (2004) 035206.

\bm{tit00}
A.I. Titov, B. K\"ampfer, and B.L. Reznik, Eur. Phys. J. A
{\bf 7} (2000) 543.

\bm{noy72}
H.P. Noyes, Annu. Rev. Nucl. Sci. {\bf 22} (1972) 465.

\bm{dmit86}
V. Dmitriev, O. Sushkov and C. Gaarde, Nucl. Phys.A {\bf 459}
(1986) 503.

\bm{risk93}
T.S.H. Lee and D.O. Riska, Phys.Rev.Lett,  {\bf 70} (1993) 1137.

\bm{horo94}
C.J. Horowitz, H.O. Meyer and D.K. Griger, Phys. Rev.C {\bf 49} (1994)
1337.

\bm{fru84}
S. Frullani and J. Mougey, Adv. Nucl. Phys. {\bf 14} (1984) 1.

\bm{shy04}
R. Shyam, H. Lenske and U. Mosel, Phys. Rev. C {\bf 69} (2004) 065205.
                                                                                
\bm{shy05}
R. Shyam, H. Lenske and U. Mosel, LANL preprint nucl-th/0505043.

\bibitem{muh04}
P. M\"uhlich, T. Falter and U. Mosel, Eur. Phys. J. A{\bf 20} (2004) 499.

\bibitem{dmi85}
V. F. Dmitriev and Toru Suzuki, Nucl. Phys. {\bf A438} (1985) 697.
 


\end{thebibliography}
\end{document}